# Chaotic saddles and interior crises in a dissipative nontwist system

R. Simile Baroni*[1], R. Egydio de Carvalho*[2], I. L. Caldas[&3], R. L. Viana[#&4], P. J. Morrison[+5]

*Universidade Estadual Paulista–UNESP, Instituto de Geociências e Ciências Exatas–IGCE, Departamento de Estatística, Matemática Aplicada e Ciências da Computação, 13506-900 Rio Claro-SP, Brazil

&Universidade de São Paulo-USP, Instituto de Física-IF, 05508-900 São Paulo-SP, Brazil

#Universidade Federal do Paraná-UFPR, Departamento de Física-DF, 80060-000 Curitiba, PR, Brazil

+ University of Texas at Austin, Institute for Fusion Studies, Department of Physics, Austin, TX 78712, USA

[1]r.baroni@unesp.br, [2]ricardo.egydio@unesp.br, [3]ibere@if.usp.br, [4]viana@fisica.ufpr.br, [5]morrison@physics.utexas.edu

## Abstract

We consider a dissipative version of the standard nontwist map. Nontwist systems present a robust transport barrier, called the shearless curve, that becomes the shearless attractor when dissipation is introduced. This attractor can be regular or chaotic depending on the control parameters. Chaotic attractors can undergo sudden and qualitative changes as a parameter is varied. These changes are called crises, and at an interior crisis the attractor suddenly expands. Chaotic saddles are nonattracting chaotic sets that play a fundamental role in the dynamics of nonlinear systems, they are responsible for chaotic transients, fractal basin boundaries, chaotic scattering and they mediate interior crises. In this work we discuss the creation of chaotic saddles in a dissipative nontwist system and the interior crises they generate. We show how the presence of two saddles increase the transient times and analyze the phenomenon of crisis induced intermittency.

**I. INTRODUCTION**

Nontwist systems naturally arise in the description of several physical phenomena, such as atmospheric zonal flows [1,2], the modeling of particle transport and magnetic field lines in plasma physics [3–7] and the motion of satellites near an oblate planet [8]. The standard nontwist map (SNM) is a discrete-time two-dimensional dynamical system that presents the universal features of nontwist dynamics [2,9–12]: isochronous resonances, separatrix and manifold reconnection, collision of periodic orbits, meandering tori and the shearless curve, which is a robust transport barrier in the phase space.

Nontwist maps are usually derived from a Hamiltonian formulation. Hamiltonian systems conserve energy and preserve the symplectic two-form [13]. Therefore, the corresponding discrete time maps are area-preserving. When a perturbation breaks the integrability of the system chaotic behavior is a possibility, along with periodic and quasiperiodic behaviors. The Poincaré-Birkhoff and KAM (Kolmogorov, Arnold, Moser) theorems describe the dynamics of the regular solutions of near-integrable systems [14]. One necessary condition for the validity of these theorems is the twist condition, that needs to be satisfied globally. The Hamiltonian of a near-integrable system with one and a half degree of freedom can be written as

$$H = H_0(J) + \epsilon H_1(\theta, J, t), \qquad (1)$$

where $(J, \theta)$ is the pair of action-angle variables, $\epsilon \ll 1$ is the perturbation parameter and $t$ is the time variable. The unperturbed frequency of the system is $\omega(J) = \partial H_0/\partial J$, and the twist condition follows as $\partial \omega(J)/\partial J \neq 0$, guaranteeing a monotonic behavior of the frequency. The discrete time analogue of the twist condition is $\partial y_{n+1}/\partial x_{n+1} \neq 0$, where $(x, y)$ are the coordinates of the map. If the twist condition is violated at a point, the frequency profile is no longer monotonic and presents an extremum, the system is called nontwist [7] and the forementioned properties are observed. The shearless curve appears at the point the twist condition is violated.

It is of natural interest to consider the effects of dissipation in nontwist systems, as real experiments often present at least a small amount of dissipation [15]. There are different ways to introduce dissipation to Hamiltonian systems [16–19]. In particular we are interested in the class of dissipative systems known as conformally symplectic [19,20], which has the property that the symplectic form is transformed into a multiple of itself, resulting in contraction of areas in the phase space. This type of dissipation models simple mechanical systems with friction proportional to the velocity, and recently conformally symplectic systems have been used to develop optimization algorithms in machine learning [21].

When dissipation is introduced, area-contractions makes it so any initial condition converges to asymptotic states, attractors, which can be regular or chaotic. The effects of dissipation on the shearless curve have been a topic of interest since first considered in the labyrinthic SNM [22–24]. The dissipation gives rise to an attractor on a torus which retains some of the characteristics of the shearless curve, such as shape and rotation number, and this attractor was called the shearless attractor (SA). The SA can be quasiperiodic or chaotic, depending on the control parameters, or not be present at all. Two routes for the transitions from the quasiperiodic SA to chaotic behavior have been previously characterized [25], as well as a route for the destruction and reappearance of the quasiperiodic SA [26].

Chaotic saddles are invariant nonattracting chaotic sets that are of fundamental importance in nonlinear systems [27–29]. They are responsible for phenomena widely observed in dynamical systems, such as chaotic transients [30,31], chaotic scattering [32,33] and fractal basin boundaries [27,34–36]. Chaotic saddles are also related to sudden and qualitative changes that chaotic attractors undergo as a system parameter is varied. Those changes are called *crises* [30,37], and the two most common types are (i) boundary crisis, which occurs when a chaotic attractor collides with an unstable periodic orbit (or equivalently its stable manifold) on the basin boundary and is converted into a nonattracting chaotic set, and (ii) interior crisis, which occurs when a chaotic attractor, typically small, collides with a chaotic saddle (or its stable manifold), resulting in a sudden enlargement of the attractor. Interior crises are also one of the mechanisms that generate intermittent behavior in dynamical systems [38]. A trajectory on the post-critical attractor spends most of the time on the small region corresponding to where the pre-crisis attractor used to be and does occasional intermittent excursions to far-away regions.

In this work we consider the dissipative SNM and present a numerical study of two qualitatively different chaotic saddles of the system, and the interior crises they cause. In Sec. II we describe the system and analyze its fixed points. In Sec. III we discuss the creation of the chaotic saddles, which we have called local chaotic saddle and global chaotic saddle. In Sec. IV we analyze the transient times of the system, and show that when both saddles are present, the transients are much longer. In Sec. V we discuss the intermittent behavior induced by the interior crisis caused by the global chaotic saddle. The paper is concluded in Sec. VI. The numerical methods are presented in the appendix.

## II. DISSIPATIVE STANDARD NONTWIST MAP

We consider the dissipative standard nontwist map (DSNM) defined by the equations:

$$M: \begin{aligned} y_{n+1} &= (1-\gamma)y_n - b\sin(2\pi x_n), \\ x_{n+1} &= x_n + a(1 - y_{n+1}^2), \quad \mod(1) \end{aligned} \quad (2)$$

where $(x_n, y_n)$ is the $n$-th iterate of a pair of dynamical variables. The parameter $b$ controls the nonlinear perturbation of the system and $a$ is related to the unperturbed rotation number profile. The dissipation is controlled by the parameter $\gamma$; if $\gamma = 0$ the area-preserving standard nontwist map of [2] is recovered, and for $0 < \gamma < 1$ the map is area-contracting which gives rise to attractors in the phase space. One attractor of particular interest is the SA, whose ancestor is the shearless torus of the conservative map that was shown in [39] to be resilient under perturbation. The shearless torus is a robust transport barrier, it divides regions of the phase space and supports the effect perturbations that tend to destroy other invariant curves. Similarly, the SA is also observed as a robust attractor, since it survives under generic perturbations and different intensities of dissipation [24].

The DSNM can present multistability in the phase space, that is, multiple attractors can coexist, and different initial conditions can converge to different attractors. The set of initial conditions that converge to an attractor defines its basin of attraction. In FIG. 1, we see the attractors and basins of attraction of the system defined by Eq. (2) for different values of $a$ and $b$. The dissipation parameter is $\gamma = 0.1$ in all of these plots. In panel (a) we have the coexistence of the quasiperiodic shearless attractor and two period-1 attractors. The basins boundaries are smooth. In panel (b) we have again the quasiperiodic shearless attractor and two period-1 attractors, but they coexist with two period-2 attractors and two period-3 attractors. Now the basin boundaries are fractalized, which is a sign that there is, at least, a chaotic saddle in the system. In panel (c) the (SA) is chaotic, it coexists with two period-4 attractors and the basin boundaries are again fractalized.

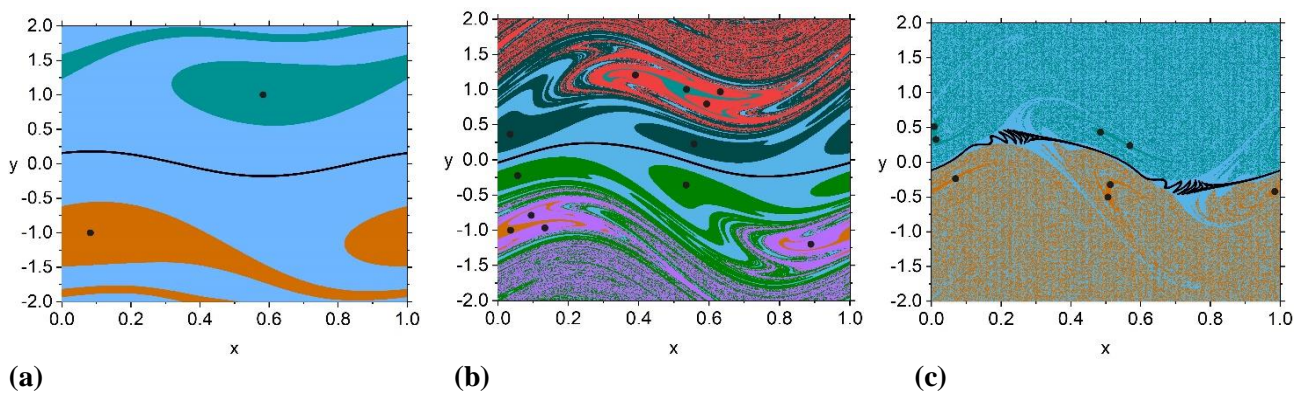

(a) (b) (c)

**FIG. 1** *(Color online). Attractors (black) and corresponding basins of attraction of the DSNM with $\gamma = 0.1$ and (a) $a = b = 0.2$, (b) $a = 0.55$ and $b = 0.45$, (c) $a = 0.59$ and $b = 0.7$.*

The fixed points $(x^*, y^*)$ of the DSNM satisfy,

$$y^* = (1-\gamma)y^* - b\sin(2\pi x^*), \qquad (3)$$

$$m = a\left(1 - y^{*2}\right), \qquad (4)$$

where $m$ is an integer number. For each $m \in \mathbb{Z}$ we have the following fixed points,

$$P_{\pm,m} = (y^*, x^*) = \left(\pm\sqrt{1 - \frac{m}{a}}, -\frac{1}{2\pi}\sin^{-1}\left(\pm\frac{\gamma}{b}\sqrt{1 - \frac{m}{a}}\right)\right). \qquad (5)$$

Two other fixed points can be found for each $m$ by noting that the map has discrete symmetries [9,39]; in particular,

$$S: \begin{aligned} y_{n+1} &= -y_n, \\ x_{n+1} &= x_n + \frac{1}{2}, \end{aligned} \qquad (6)$$

i.e., $S \circ M = M \circ S$. Applying the symmetry transformation to Eq. (5) we find the fixed points,

$$SP_{\pm,m} = (y^*, x^*) = \left( \mp\sqrt{1 - \frac{m}{a}}, -\frac{1}{2\pi}\sin^{-1}\left(\pm\frac{\gamma}{b}\sqrt{1-\frac{m}{a}}\right) + \frac{1}{2} \right). \tag{7}$$

The stability of a fixed point can be assessed by checking the eigenvalues of the Jacobian matrix of the map $M$, evaluated at the fixed point. The Jacobian matrix is

$$J = \begin{pmatrix} J_{11} & J_{12} \\ J_{21} & J_{22} \end{pmatrix}, \tag{8}$$

whose elements are given by the derivatives,

$$J_{11} = \frac{\partial y_{n+1}}{\partial y_n} = 1 - \gamma, \tag{9}$$

$$J_{12} = \frac{\partial y_{n+1}}{\partial x_n} = -2\pi b \cos(2\pi x_n), \tag{10}$$

$$J_{21} = \frac{\partial x_{n+1}}{\partial y_n} = -2a(1-\gamma)[(1-\gamma)y_n - b\sin(2\pi x_n)], \tag{11}$$

$$J_{22} = \frac{\partial x_{n+1}}{\partial x_n} = 1 + 4\pi ab \cos(2\pi x_n)[(1-\gamma)y_n - b\sin(2\pi x_n)], \tag{12}$$

and the eigenvalues of $J$ are generically given by,

$$\lambda_{\pm}(x_n, y_n; a, b, \gamma) = \frac{\text{Tr}(J)}{2} \pm \frac{1}{2}\sqrt{\text{Tr}(J)^2 - 4\,\text{Det}(J)}, \tag{13}$$

where $\text{Tr}(J) = J_{11} + J_{22}$ is the trace and $\text{Det}(J) = J_{11}J_{22} - J_{12}J_{21}$ is the determinant of $J$. As the expression in the right-hand-side of Eq. (13) is very large it is not written explicitly here, but we can easily get that $\text{Det}\,J = 1 - \gamma$, so that for $\gamma = 0$ the map is area-preserving and the phase space is mixed, with chaotic seas and regular structures. On the other hand, for $0 < \gamma < 1$, $\text{Det}\,J < 1$, which implies that the map contracts areas and there are attractors in the phase space.

In the dissipative configuration, there are two possibilities for the nature of the fixed points. If the eigenvalues of $J$ calculated at the fixed point are a complex conjugate pair, the fixed point is a stable focus, which is a period-1 point attractor. The other possibility occurs when the eigenvalues are real, with $\lambda_+ \lessgtr 1 \lessgtr \lambda_-$, which gives rise to a saddle point, that is, a period-1 unstable periodic orbit (UPO).

Evaluating $\lambda_\pm$ at the fixed points $P_{+,m}$ and $SP_{+,m}$ we find

$$\lambda_\pm(P_{+,m}; a, b, \gamma) = \lambda_\pm(SP_{+,m}; a, b, \gamma) \qquad (14)$$

$$= \frac{1}{2}\left(2 - \gamma + 4\pi b \sqrt{a(a-m)\left(1 + \frac{\gamma^2}{ab^2}(m-a)\right)}\right.$$

$$\left. \pm \sqrt{4(\gamma - 1) + \left(-2 + \gamma - 4\pi b \sqrt{a(a-m)\left(1 + \frac{\gamma^2}{ab^2}(m-a)\right)}\right)^2}\right),$$

so that the pair $P_{+,m}$ and $SP_{+,m}$, for a given set of control parameters, always have the same stability. The same is true for the pair $P_{-,m}$ and $SP_{-,m}$, for which we find

$$\lambda_\pm(P_{-,m}; a, b, \gamma) = \lambda_\pm(SP_{-,m}; a, b, \gamma) \qquad (15)$$

$$= \frac{1}{2}\left(2 - \gamma - 4\pi b \sqrt{a(a-m)\left(1 + \frac{\gamma^2}{ab^2}(m-a)\right)}\right.$$

$$\left. \pm \sqrt{4(\gamma - 1) + \left(-2 + \gamma + 4\pi b \sqrt{a(a-m)\left(1 + \frac{\gamma^2}{ab^2}(m-a)\right)}\right)^2}\right).$$

When a fixed point's stability changes as a control parameter is varied, or new fixed points appear, it is said that a bifurcation has happened. Specifically, when an eigenvalue crosses the complex unit circle at $\lambda = +1$ a *saddle-node bifurcation (SN)* happens, that is, a period-1 UPO and a period-1 point-attractor are created or destroyed. When the crossing happens at $\lambda = -1$, a *period-doubling bifurcation (PD)* happens. For more details, see reference [40].

We use Eqs. (14) and (15) to find parameter values for bifurcations, by setting the eigenvalues equal to $+1$ or $-1$. Then we solve for a parameter as a function of the others, and the resulting expression gives us the surface in the parameter space where a bifurcation happens [40].

Following this procedure, we find that for a given $m$, all four fixed points are created via a saddle-node bifurcation at the same parameter values. Solving either $\lambda_\pm(P_{+,m}; a, b, \gamma) = +1$ or $\lambda_\pm(P_{-,m}; a, b, \gamma) = +1$ for $b$, we find,

$$b_{SN,m}(a,\gamma) = \pm\gamma\sqrt{1-\frac{m}{a}}, \quad (16)$$

and doing the same with $\lambda_{\pm}(P_{+,m}; a,b,\gamma) = -1$ or $\lambda_{\pm}(P_{-,m}; a,b,\gamma) = -1$ we get,

$$b_{PD,m}(a,\gamma) = \pm\frac{1}{2\pi\sqrt{a(m-a)}}\sqrt{-4 + 4\gamma - \gamma^2(1 + 4a^2\pi^2 - 8am\pi^2 + 4m^2\pi^2)}. \quad (17)$$

Fixing $\gamma$, for a given value of $m$, Eq. (16) defines a curve in the $(a,b)$ parameter space where the fixed points are born. Similarly, Eq. (17) defines a curve where the fixed points go through period-doubling. In this work we consider only positive values of $b$, so the equations with the minus sign are discarded.

In FIG. 2 we show some period-doubling and saddle-node bifurcation curves in the $(a,b)$ parameter space with $\gamma = 0.1$. The solid lines are saddle-node bifurcation curves while the dashed lines correspond to period-doubling bifurcations.

The bifurcations of the $m = 0$ fixed points are represented in black. For this value of $m$, the saddle-node bifurcation curve does not depend on $a$ and is a straight line at a constant value of $b$, as we get $b_{SN,m=0}(a,\gamma) = \gamma$ from setting $m = 0$ in Eq. (16). We see that the two period-doubling bifurcation curves are symmetric with respect to the $a = 0$ axis, as are the positive and negative segments of the saddle-node bifurcation curve.

For $m = 1$, Eq. (17) results in,

$$b_{SN,m=1}(a,\gamma) = \gamma\sqrt{1-\frac{1}{a}}. \quad (18)$$

For positive values of $a$, this expression is real valued for all $a \geq 1$ and assymptotically approaches $b_{SN,m=1}(a,\gamma) = \gamma$ as $a \to \infty$. Eq. (18) is also real for all $a < 0$, it also approaches $b_{SN,m=1}(a,\gamma) = \gamma$ as $a \to -\infty$ and $b_{SN,m=1}(a,\gamma) \to -\infty$ as $a \to 0$. We see that the bifurcation curves in the case $m = 1$, are no longer symmetric with respect to the $a = 0$ axis, as was the case for $m = 0$. However, the $m = -1$ bifurcation curves are the reflection of the $m = 1$ bifurcation curves with respect to the $a = 0$ axis.

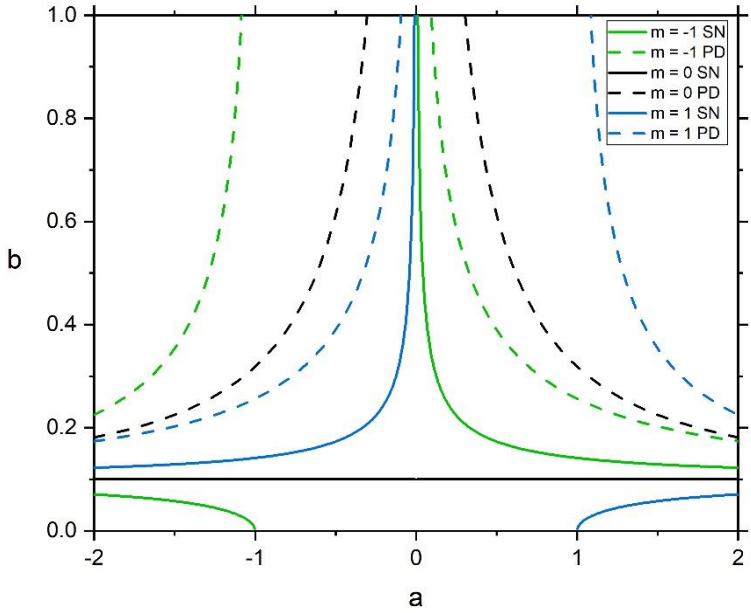

**FIG. 2** *(Color online). Saddle-node and period-doubling bifurcation curves of the $m = -1, 0, +1$ families of fixed points of the DSNM, with $\gamma = 0.1$.*

In this work we are particularly interested in positive values of both $a$ and $b$. In FIG. 3 we again show the bifurcation curves for a few families of fixed points, but we restrict the plot to the $a, b \in (0,1]$ quadrant of the parameter space and include more values of $m$.

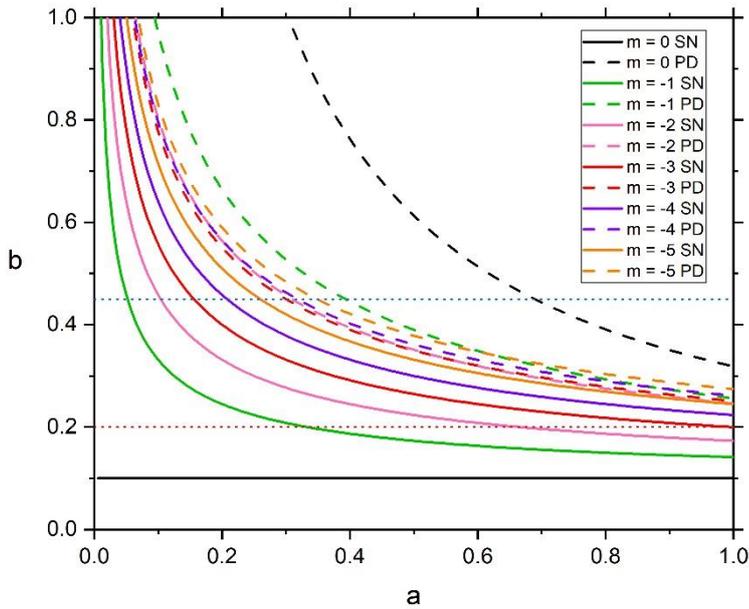

**FIG. 3** *(Color online). Saddle-node and period-doubling bifurcation curves of the $m = -5$ to $m = 0$ families of fixed points of the DSNM, with $\gamma = 0.1$. The dotted red and blue lines are $b = 0.2$ and $b = 0.45$.*

To visualize the fixed points of the system and the changes in their stability as $a$ is varied, we plot the $y$ coordinate of each of them, given by Eqs. (5) and (7), as long as $x$ and $y$ are real valued. We distinguish stable and unstable via Eqs. (14) and (15), when the fixed point is stable it is colored blue, and when it is unstable it is colored red. The stability diagram of $P_\pm$ and $SP_\pm$ are shown in FIG. 4 in panels (a) and (b), respectively, for $b = 0.2$.

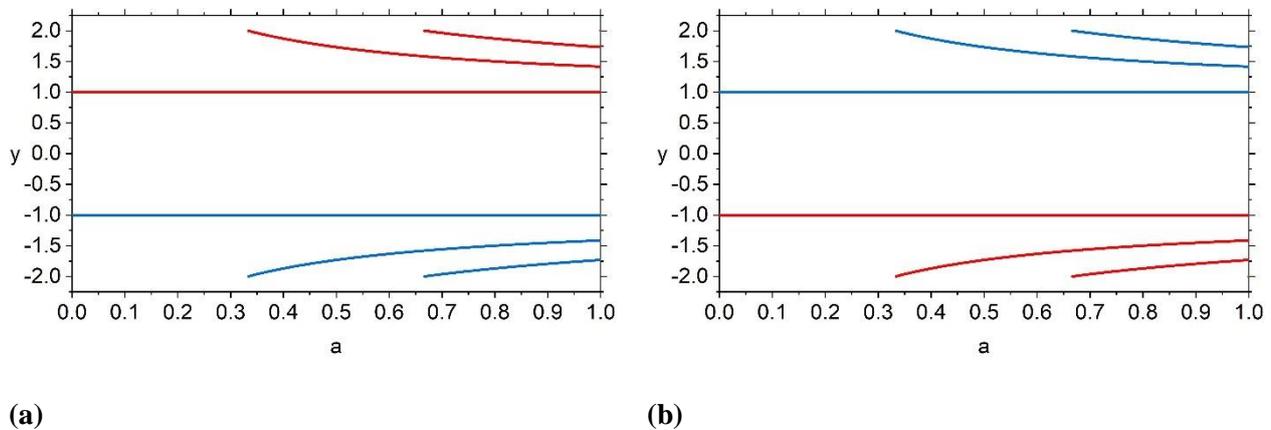

(a)　　　　　　　　　　　　　　　　　　(b)

**FIG. 4**  *(Color online). Fixed points stability diagram for $\gamma = 0.1$ and $b = 0.2$. Blue represents stability and red instability. (a) $P_\pm$, (b) $SP_\pm$.*

Analyzing FIG. 4 we can see the creation of the $m = -1$ fixed points around $a = 0.35$. This agrees with FIG. 3, where we can see the $b = 0.2$ dotted red line crossing the $m = -1$ saddle-node bifurcation curve, drawn in green, around that value of $a$. The birth of the $m = -2$ fixed points, in FIG. 3, when $a$ crosses the respective saddle-node bifurcation curve, can also be seen around $a = 0.7$. We also see that $P_{+,m}$ and $SP_{+,m}$ are always unstable for the considered parameters, while $P_{-,m}$ and $SP_{-,m}$ are always stable.

In FIG. 5 we present the stability diagrams for a higher perturbation value, $b = 0.45$. We now see the creation of many more fixed points, reaching $m = -20$. We can also see stable fixed points becoming unstable, this happens after the period-doubling bifurcations. For instance, $P_{-,m=0}$ and $SP_{-,m=0}$ become unstable around $a = 0.7$, agreeing with the result shown in FIG. 3, where we can see the $b = 0.45$ dotted blue line crossing the $m = 0$ period-doubling bifurcation curve.

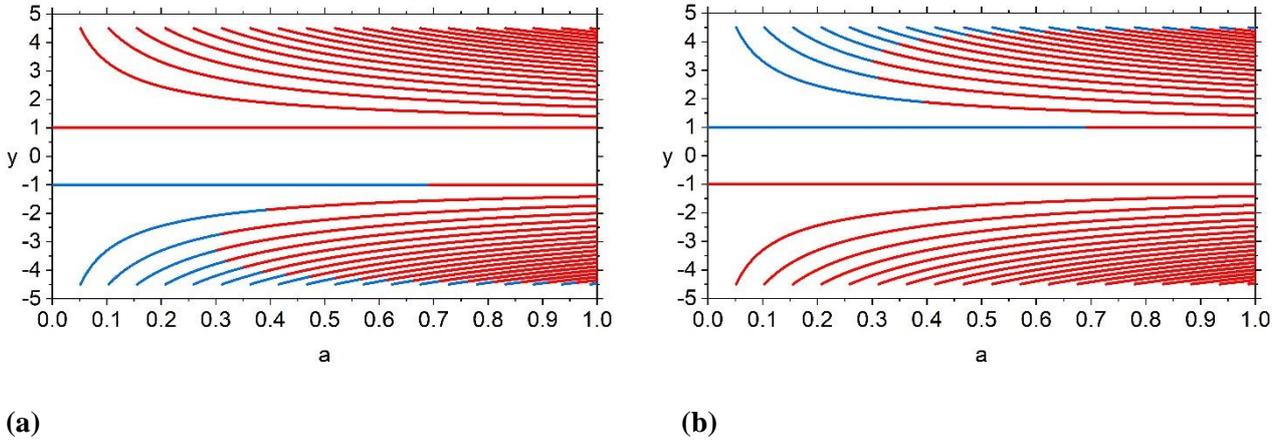

**FIG. 5**  *(Color online). Fixed points stability diagram for $\gamma = 0.1$ and $b = 0.45$. Blue represents stability and red instability. (a) $P_\pm$, (b) $SP_\pm$.*

The higher the value of $|m|$, the narrower the parameter range for which the stable fixed points remain stable. The higher the value of $b$, the more families of fixed points are created as $a$ is varied. We also note that for a fixed $b$ all fixed points with $m < 0$ are created at $y = \pm 10\,b$, and $y$ decreases as $a$ is increased. In FIG. 4 we have $b = 0.2$ and we see the $m = -1$ and $m = -2$ fixed points are created at $y = 2$. In FIG. 5, for $b = 0.45$, we see fixed points from $m = -20$ to $m = -1$ being created at $y = 4.5$ as $a$ is varied.

Although this bifurcation analysis has been carried only for the fixed points of the map, periodic orbits with higher periods are also found in the system, and they exhibit similar bifurcations. However, it is not possible to find their coordinates or bifurcation curves analytically.

## III. CREATION OF CHAOTIC SADDLES

Two qualitatively different types of chaotic saddles are observed in the DSNM, we call them *global chaotic saddle* (GCS) and *local chaotic saddle (LCS)*. In this section we describe how these non-attracting chaotic sets are created.

The (GCS) is a two-piece chaotic saddle that exists throughout a larger range of control parameters, when compared to the (LCS). One of its pieces occupies the upper half of the phase space, while the other one is in the lower half, and both are separated by the shearless attractor. This saddle is responsible for the interior crisis discussed in Sec. V.

To understand how the (GCS) is created, we fix $\gamma = 0.1$ and choose a value of perturbation $b$. We wish to simultaneously see all the attractors of the system and their bifurcations as $a$ is varied from 0 to 1. To do so we start with $a = 0$ and distribute 75 initial conditions over a straight line that crosses the phase space from $(x, y) = (0, -10)$ to $(x, y) = (1, 10)$. We iterate each one $10^4$ times and save the last 10 iterations of the $y$ coordinate. Then we move on to the next value of $a$ and the same initial conditions are iterated again. The process is repeated until it reaches $a = 1$. The resulting plot of the last iterations of $y$ coordinates versus $a$ is a global bifurcation diagram, and it is shown in FIG. 6(a) for $b = 0.2$ and FIG. 6(b) for $b = 0.45$.

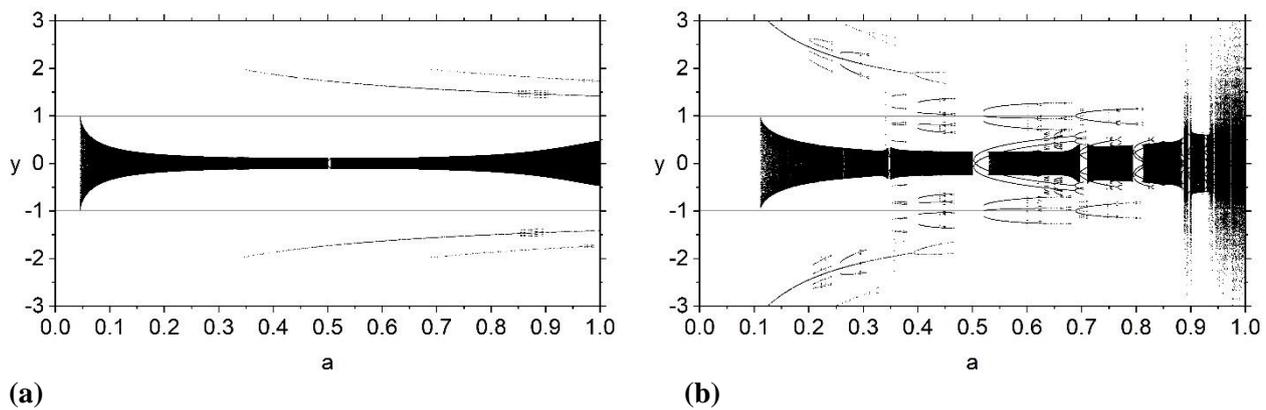

**FIG. 6** *Global bifurcation diagram for $\gamma = 0.1$ and (a) $b = 0.2$, (b) $b = 0.45$.*

Analyzing FIG. 6(a) we can see all the stable fixed points depicted in FIG. 4. Additionally, the shearless attractor appears in the $|y| < 1$ region. We can also see some higher period periodic attractors around $a = 0.85$ near the $m = -1$ point attractors.

In FIG. 6(b), where we have a higher value of perturbation, $b = 0.45$, we see an increased number of point attractors, of multiple periods, appearing via saddle-node bifurcations. The $m = 0$, $m = -1$ and $m = -2$ stable fixed points are visible, they suffer period-doubling bifurcations and become unstable agreeing with the results in FIG. 3 and FIG. 5. We also note an abrupt increase in the size of an attractor of the system around $a = 0.9$. This event is an *interior crisis*.

In the global bifurcation diagrams of FIG. 6, some point attractors appear as dotted lines, which gives the impression that they are appearing and disappearing as $a$ is varied. However, this behavior is because their basins of attraction are very small when compared to the other attractors', so when computing the diagrams, it may happen that the initial conditions do not fall into their basins.

With each saddle-node bifurcation, new UPOs are created. The homoclinic and heteroclinic crossings of their stable and unstable manifolds generate the global chaotic saddle, allowing phenomena such as transient chaos and fractal basin boundaries [36].

The second type of chaotic saddle observed in the system only exists within small ranges of parameters, inside periodic windows. Therefore, we call them local chaotic saddles. A periodic window begins when a chaotic attractor loses stability because of a saddle-node bifurcation and is replaced by a periodic attractor. Some of the infinite number of UPOs that used to be a part of the attractor becomes the local chaotic saddle, which dramatically interferes in the transient times as will be discussed in Sec. IV.

To illustrate the creation of a local chaotic saddle, we investigate the periodic window shown in the bifurcation diagram of FIG. 7.

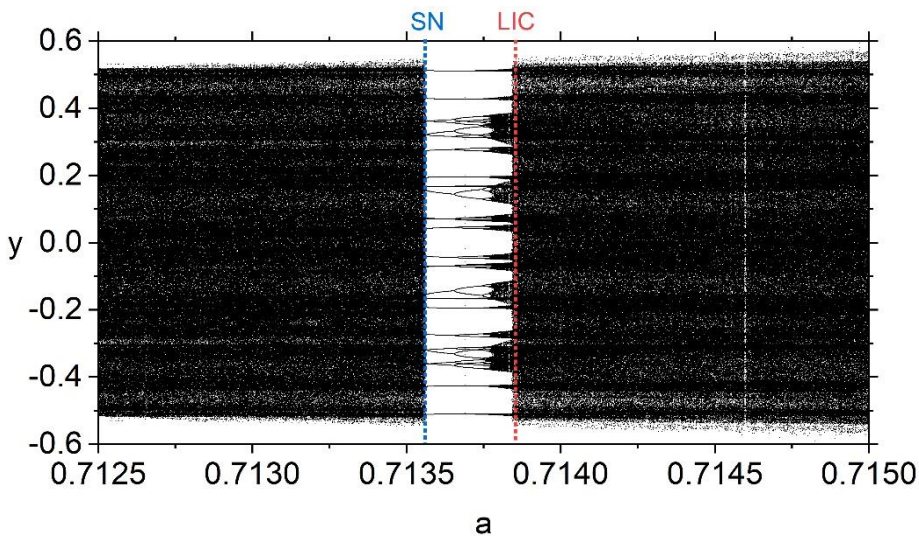

**FIG. 7** *(Color online). Global bifurcation diagram for $\gamma = 0.1$ and $b = 0.58$. The dotted blue line indicates the start of the period 26 periodic window, with a saddle-node bifurcation (SN). The dotted red line indicates the end of the periodic windows, with a local interior crisis (LIC).*

For $a = 0.7135$, before the (SN), the system has a chaotic attractor and a global chaotic saddle, shown in FIG. 8(a). The details regarding the numerical methods for finding chaotic saddles are discussed in the appendix. At the saddle-node bifurcation, a period-26 attractor is created alongside a period-26 UPO. The periodic attractor goes through a period doubling cascade creating a chaotic attractor with 26 bands that coexists with two chaotic saddles, the one that already existed before the saddle-node bifurcation and the local chaotic saddle, created at the saddle-node bifurcation. This is shown in FIG. 8(b), for $a = 0.7138$. The banded chaotic attractor collides with the local chaotic saddle, resulting in the appearance of a single banded chaotic attractor, illustrated in FIG. 8(c) for $a = 0.71385$. The growth in size marks the interior crisis, and because it happened via a collision of the chaotic attractor with the local chaotic saddle, we refer to it as a local interior crisis (LIC). In FIG. 8(d) we show an amplification of the boxed region of FIG. 8(b).

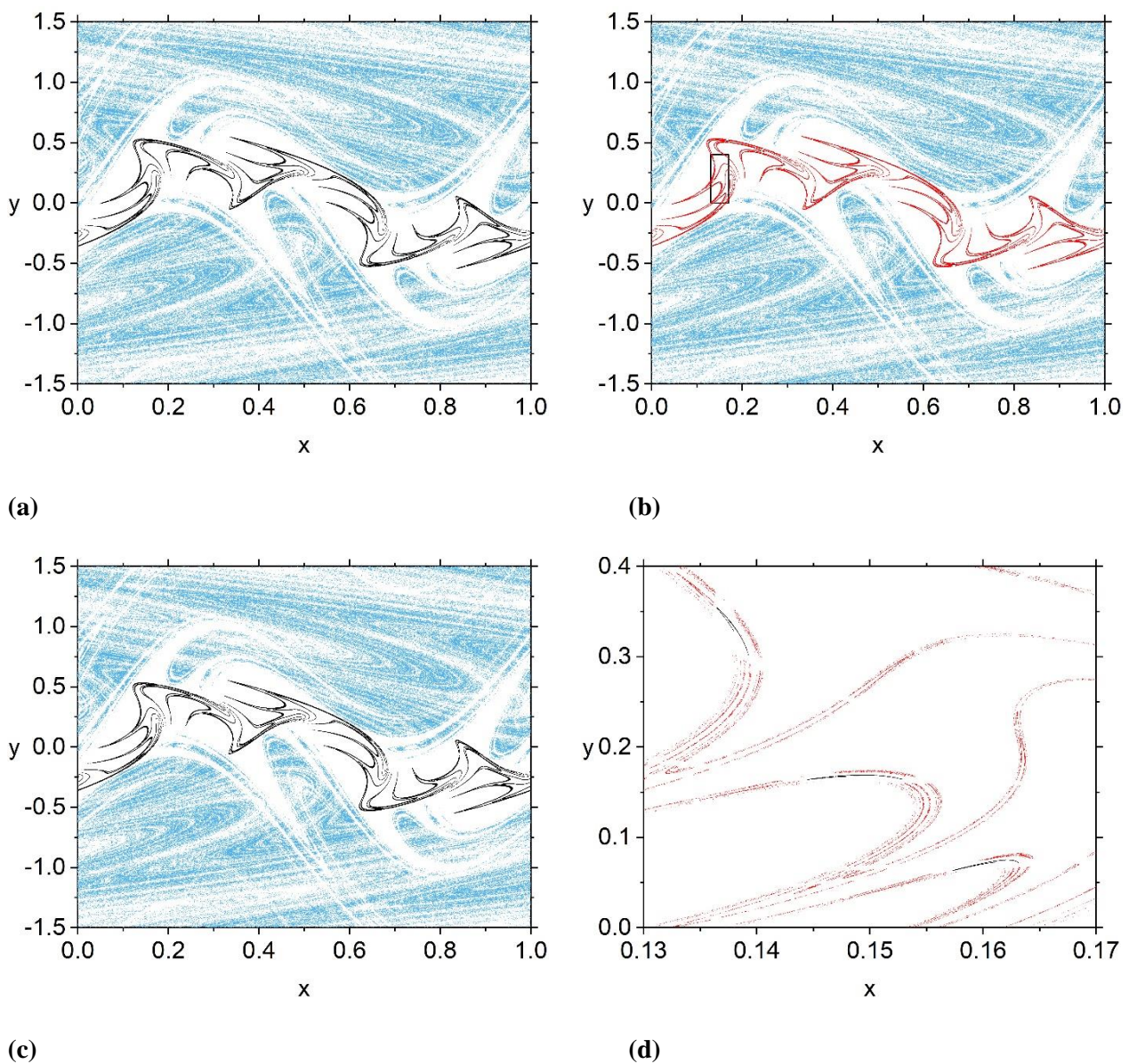

(a)  (b)  (c)  (d)

**FIG. 8** *(Color online). Attractor and chaotic saddles for $\gamma = 0.1$, $b = 0.58$ and (a) $a = 0.7135$, (b) $a = 0.7138$, (c) $a = 0.71385$, (d) $a = 0.7138$ amplified around three bands of the*

*chaotic attractor. The global chaotic saddle is colored blue, and the local chaotic saddle is colored red. The attractors are colored black.*

## IV. TRANSIENT TIMES ANALYSIS

In dynamical systems with no chaotic saddles the transient times are typically short and do not exhibit chaotic features [34]. Chaotic saddles are responsible for generally longer and chaotic transients [30,31]. Initial conditions starting close to the stable manifold of a chaotic saddle follow the saddle and spend some time in its neighborhood before leaving it by its unstable manifold. The *escape rate* is a measure how quickly this occurs. To estimate the escape rate of a chaotic saddle, when there is only one of them in the system, such as in FIG. 8(a) and FIG. 8(c), we distribute a large number of initial conditions $N_0 = 10^6$ in the phase space region defined by $0 < x < 1$ and $|y| < 1.5$, and excluding any attractor (see appendix for details). We then iterate the remaining initial conditions, which come close to the nonattracting set before leaving its neighborhood and reaching the attractor. Let $N(n)$ be the number of initial conditions that have not converged to the attractor after $n$ iterates of the map. As $n$ is increased, we observed the exponential decay $N(n)$, as reported in [27,34],

$$N(n) \sim e^{-\kappa n}, \qquad (19)$$

where $\kappa$ is the escape rate of the saddle. By this definition, $N(n)$ is decrease by a factor of $1/e$ after $1/\kappa$ iterations, meaning that most trajectories converge to the attractor after fewer than $1/\kappa$ iterations. So, we can estimate the average transient time as

$$\tau \approx \frac{1}{\kappa}. \qquad (20)$$

We start by analyzing the transient times of the system outside the periodic window, when only the global chaotic saddle is present. In FIG. 9(a) and FIG. 9(b) we estimate the escape rate of the global chaotic saddles shown in FIG. 8(a), before the periodic window, and in FIG. 8(c), after the local interior crisis. Both escape rates are comparable, reflecting the similarity of the involved chaotic saddles. We also estimate the average transient times, and we see that for both set of parameters, these quantities are about the same.

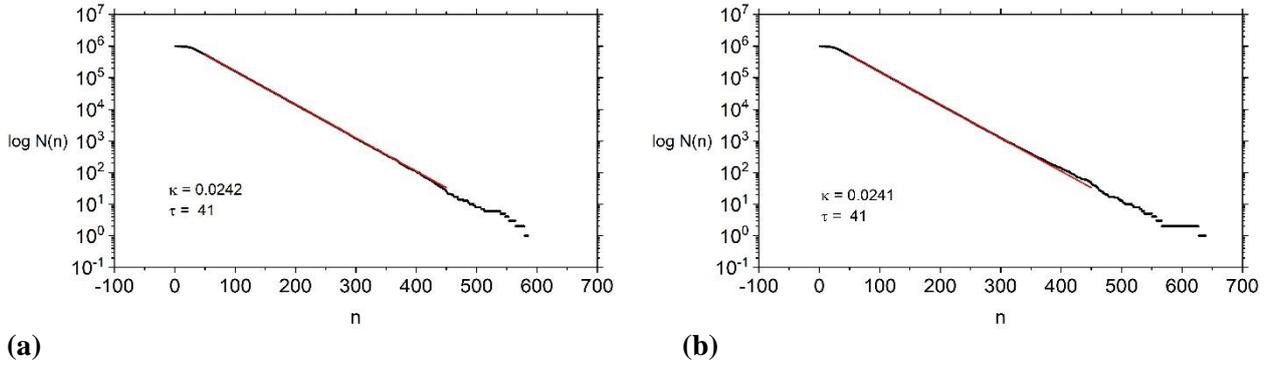

**FIG. 9**  *(Color online). Estimation of the global chaotic saddle's escape rate and average transient time. $\gamma = 0.1$, $b = 0.58$ and (a) $a = 0.7135$, before the periodic window, and (b) $a = 0.71385$ after the local interior crisis*

Another way to visualize the transient times is color-coding each initial condition according to the number of iterations required to reach the attractor. This is shown in FIG. 10(a) and FIG. 10(b), again for $a = 0.7135$, before the periodic window, and $a = 0.71385$, after the local interior crisis. The emerging intricate structure of longer transient times is reminiscent of the stable manifold of the GCS, as an initial condition closer to the stable manifold takes longer to reach the attractor.

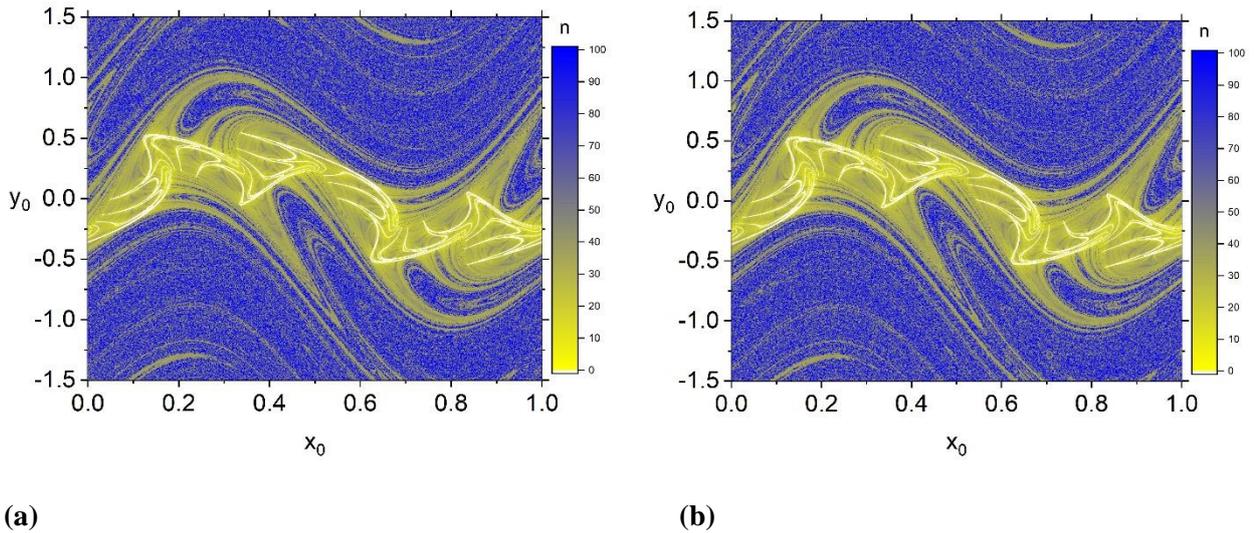

**FIG. 10**  *(Color online). Initial conditions colored according to the number of iterations needed to reach the attractor. $\gamma = 0.1$, $b = 0.58$ and (a) $a = 0.7135$, before the periodic window, and (b) $a = 0.71385$ after the local interior crisis.*

We now proceed to examine the transient behavior inside the periodic window, when both global and local chaotic saddles are present, for $a = 0.7138$. To evaluate the average transient time, we repeat the procedure carried out to evaluate the escape rate of the global saddle outside the periodic window, in FIG. 9. Note that now the system has two chaotic saddles, and so the quantity $\tau$ extracted from the exponential fitting does not represent the escape rate of either of them, but its inverse $1/\tau$ gives an estimate of the average transient time. The result is shown in FIG. 11(a), and we see that the presence of the LCS generates a longer transient. Then we repeat the same procedure but changing the escape condition. In Fig. 11(a) the escape condition was to reach the attractor, and in Fig. 11(b) the escape condition is to reach the unstable manifold of the LCS's unstable manifold, obtained with the sprinkler method (see appendix). We see that the escape rate and the average transient time obtained now (Fig. 11(b)) are the same as the ones of the GCS, obtained in Fig. 9. Therefore, the chaotic transient can be decomposed in two phases: the fast phase caused by the GCS, with average length of 41 iterations, and the slow phase caused by the LCS, with average length of 304 iterations.

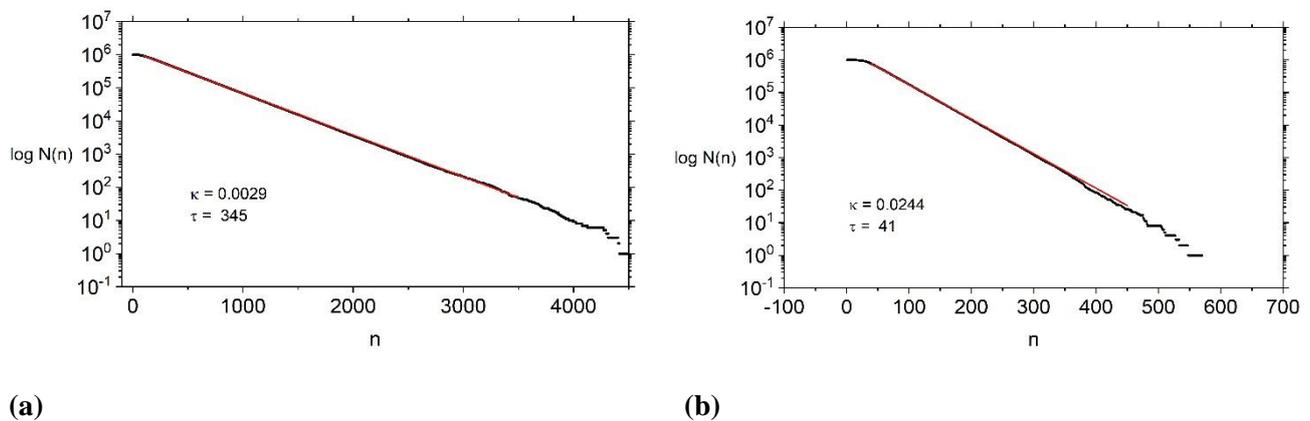

(a)  (b)

**FIG. 11** *(Color online). Estimation of escape times for $\gamma = 0.1$, $b = 0.58$ and $a = 0.7138$, inside the periodic window. (a) escape condition is reaching the attractor, and (b) escape condition is reaching the unstable manifold of the local chaotic saddle.*

In Fig. 12 we color code the initial conditions according to the necessary number of iterations to reach the attractor in panel (a), and to reach the unstable manifold of the LCS, in panel (b). We see that Fig. 12(b) is very similar to Fig. 10(a) and Fig. 10(b).

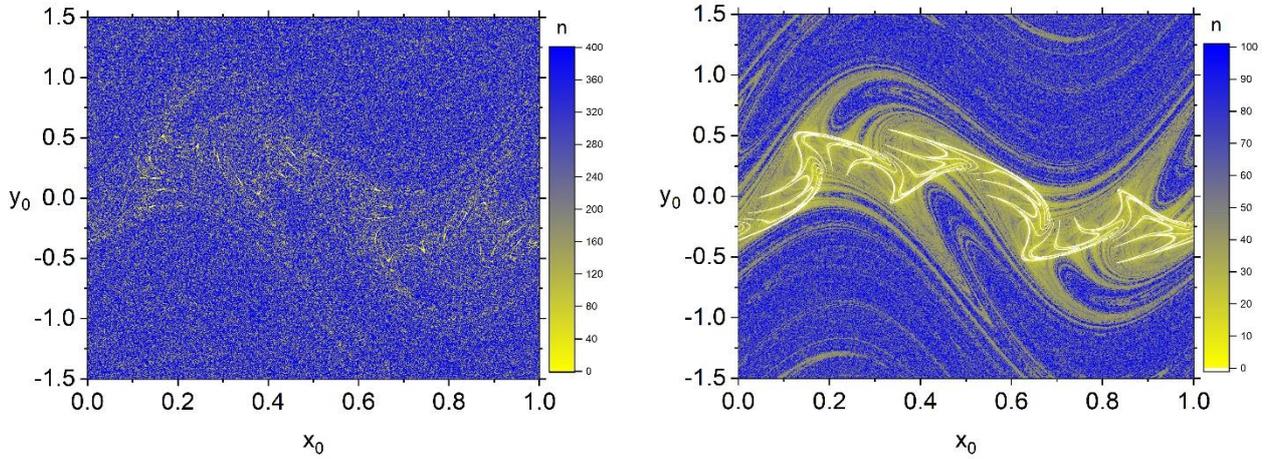

(a)       (b)

**FIG. 12**  *(Color online). Initial conditions colored according to the number of iterations needed to reach (a) the attractor and (b) the unstable manifold of the local saddle.* $\gamma = 0.1$, $b = 0.58$ *and* $a = 0.7138$.

## V. CRISIS INDUCED INTERMITTENCY

In the previous section we explored a local interior crisis, which happens when the chaotic attractor collides with the LCS and we showed how the transient times are affected by the presence of the two chaotic saddles. Now we focus on a global interior crisis, caused by the collision of the chaotic attractor with the GCS, to illustrate the phenomenon of crisis induced intermittency.

In the bifurcation diagram of FIG. 13 we illustrate a global interior crisis, with $b = 0.55$. It was computed with a single initial condition iterated $10^4$ times for each value of $a$; the $y$-coordinates of the last 5000 iterations are shown. We see that as the parameter $a$ is varied there is a sudden growth of the attractor's size, which is a consequence of the global interior crisis.

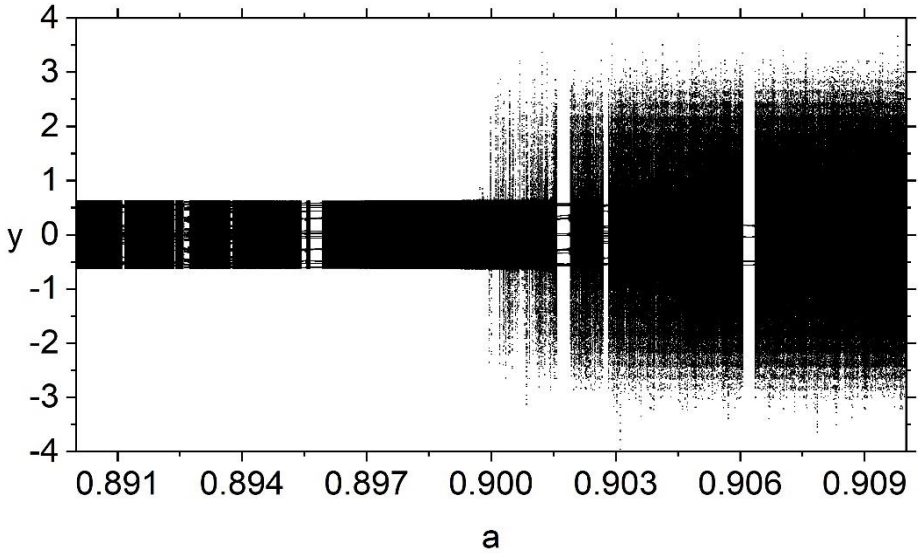

**FIG. 13** *(Color online). Bifurcation diagram for $b = 0.55$. The sudden growth of the attractor is a consequence of the global interior crisis.*

To numerically estimate the value of $a$ when the crisis happens, we do a large number of iterations ($10^7$) of an initial condition for 1000 values of $a$ between $a = 0.897$ and $a = 0.9$. We consider that the crisis has happened when a trajectory visits a region of the phase space with $|y| > 1$. With this procedure we found the critical value of $a$ to be around $a_c \cong 0.89954425$.

In FIG. 14(a) we show the chaotic attractor and the GCS for $a = 0.89$, just before the crisis. In FIG. 14(b-d) we show the post-crisis attractor for $a = 0.9$, $a = 0.901$ and $a = 0.903$, respectively. To draw the attractors, $10^6$ iterations of the map were considered. We see that in the post-crisis configurations trajectories on the attractor now visit the regions that were previously occupied by the GCS. We refer to those visits as bursts, in FIG. 15 (a-c), and they occur intermittently in time.

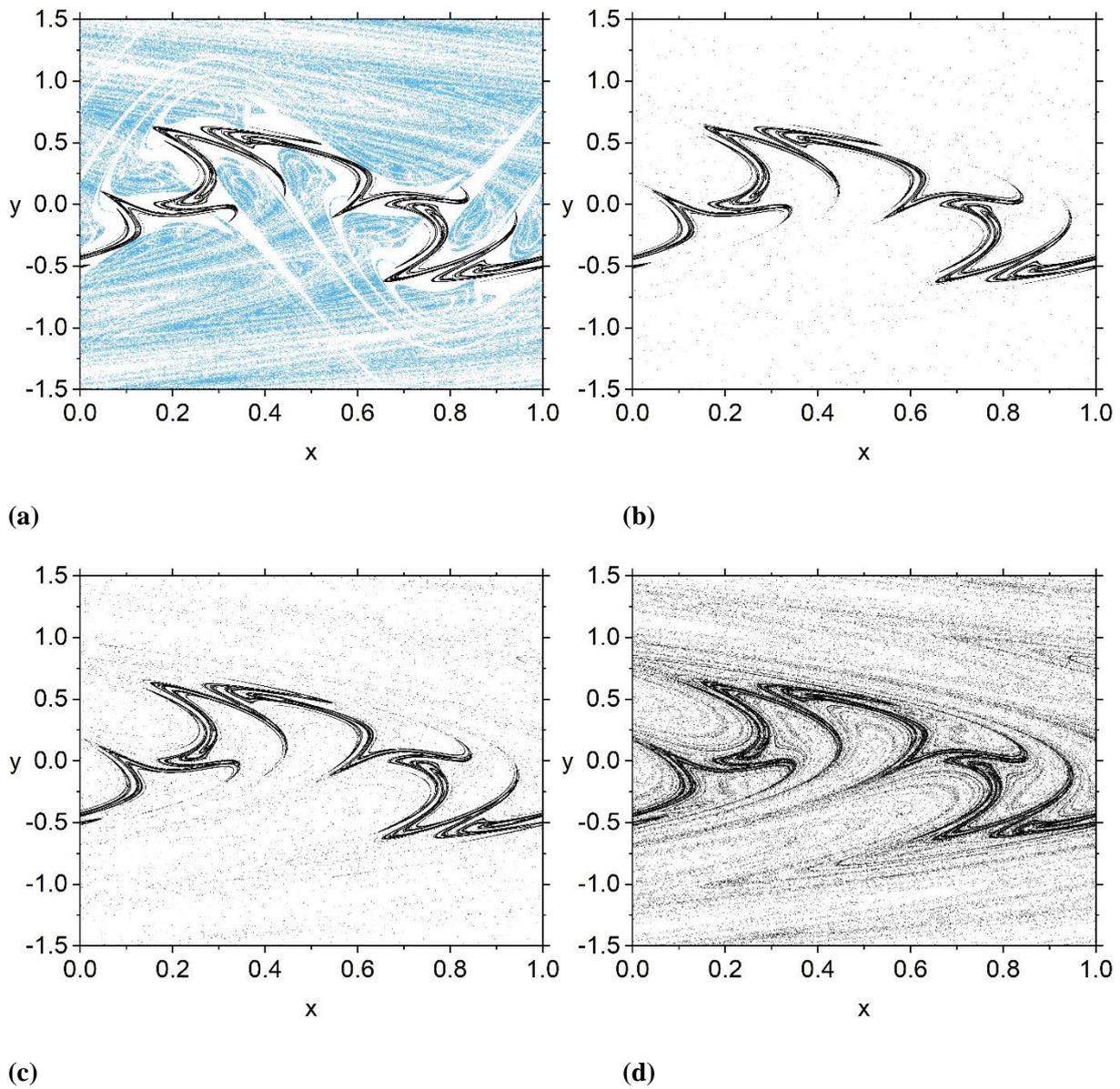

**FIG. 14** *(Color online). (a) pre-crisis attractor and GCS for $a = 0.899$, post-crisis attractor for (b) $a = 0.9$, (c) $a = 0.901$ and (d) $a = 0.903$.*

As $a$ is increased beyond the critical value $a_c$ the bursts become more frequent in time. The irregular alternation between visiting the pre-crisis region and the region previously occupied by the GCS can be visualized in a time series of the $y$-coordinate, such as those of FIG. 15.

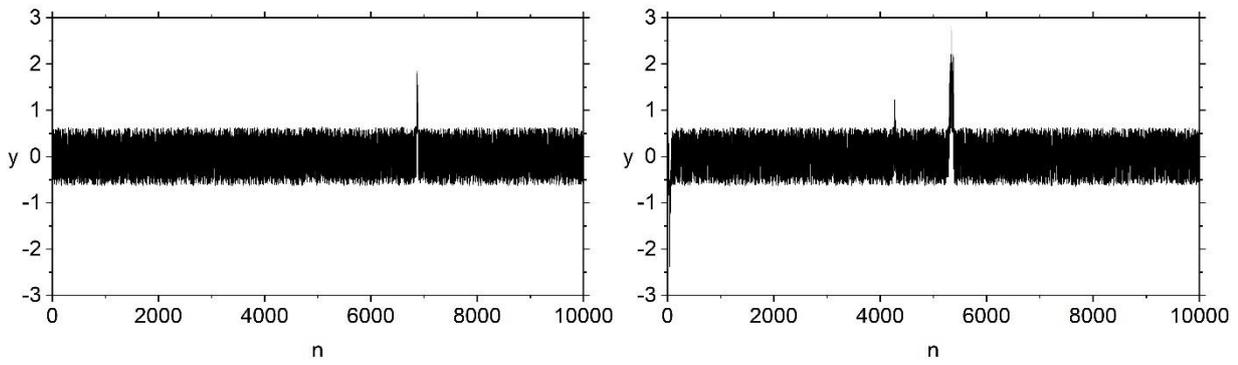
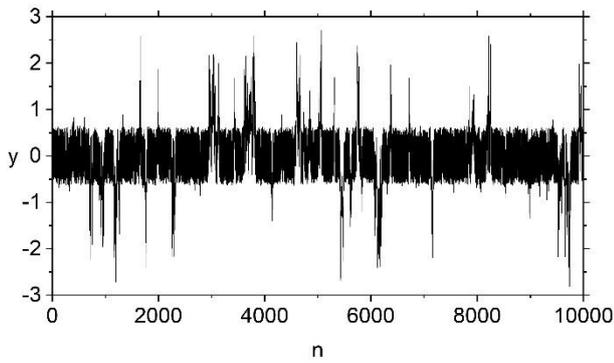

(a)

(b)

(c)

**FIG. 15**   *Time series of the y-coordinate of a trajectory on the post-crisis attractor for (a) $a = 0.9$, (b) $a = 0.901$ and (c) $a = 0.903$.*

To quantify the time between the bursts we consider that a burst starts whenever $|y| > 1$ and has ended when $|y| < 1$ for at least 50 consecutive iterations of the map. In FIG. 16 we show histograms of the time between consecutives bursts, $t$, and the average time between bursts, $\langle t \rangle$. For each value of $a$, the initial condition was iterated $10^8$ times to compute the histogram and $\langle t \rangle$.

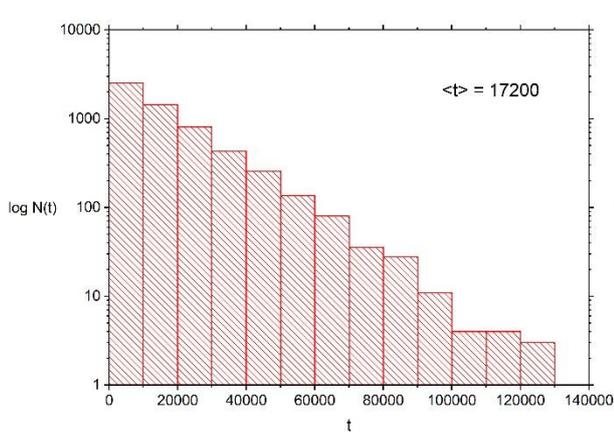
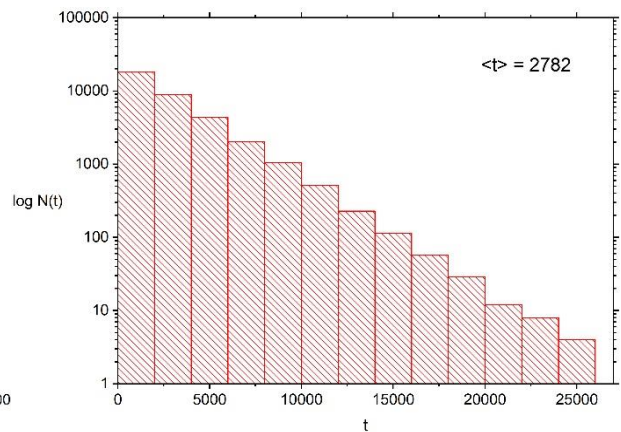

(a)

(b)

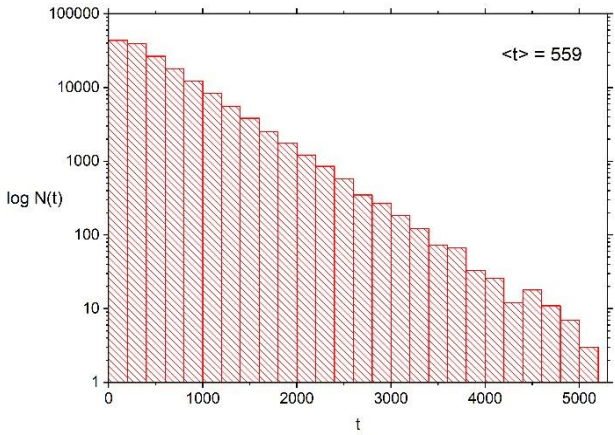

(c)

**FIG. 16** *Histogram of time between consecutive bursts $t$, and average time between consecutive bursts $\langle t \rangle$. (a) $a = 0.9$, bin size 10000, (b) $a = 0.901$, bin size 2000, and (c) $a = 0.903$, bin size 200.*

The average time between bursts is an accessible quantity to characterize the intermittent behavior. At the crisis value there is no burst and therefore $\langle t \rangle$ is infinite, and beyond the crisis the bursts become more and more frequent, so that $\langle t \rangle$ decreases as the parameter is changed. In FIG. 17 we show $\langle t \rangle$ as a function of $a$.

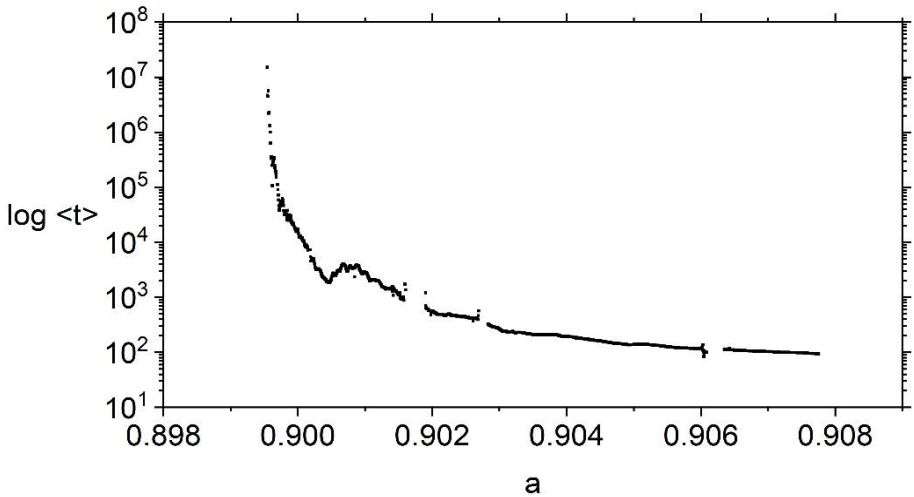

**FIG. 17** *Average time between consecutive bursts as a function of $a$.*

Crisis induced intermittency is also observed after the local interior crisis discussed in section IV, but the two alternating dynamical phases are not easily distinguishable from a time series, as is the case after the global interior crisis. However, from the natural measure (see Appendix) of the attractor after the local interior crisis, shown in FIG. 18, we note that the region previously occupied by the pre-crisis attractor is visited way more often.

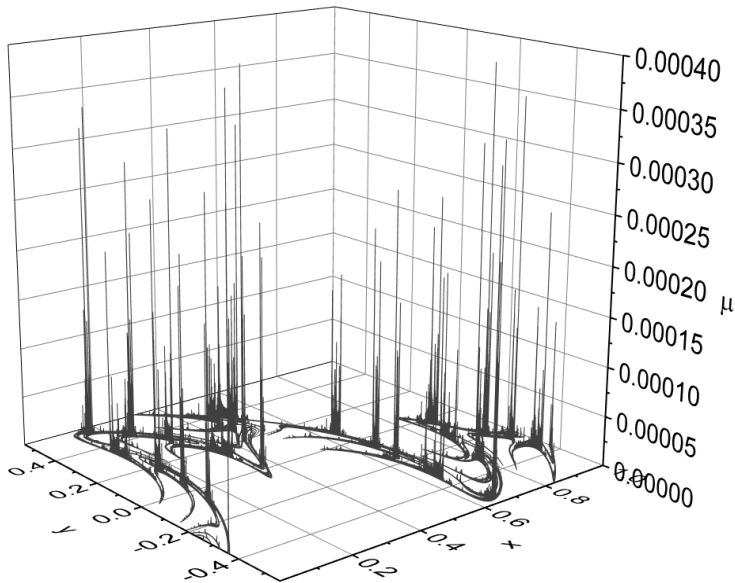

**FIG. 18**  *Natural measure $\mu$ of the chaotic attractor after the local interior crisis for $b = 0.58$ and $a = 0.71385$. The larger peaks are the regions of the attractor that were visited more frequently and correspond to the regions previously occupied by the pre-crisis attractor.*

## V. CONCLUDING REMARKS

We studied a dissipative nontwist map and we found families of fixed points. We also assessed their stability by evaluating the eigenvalues of the Jacobian matrix. From the eigenvalues expressions we calculated the saddle-node and period-doubling bifurcation curves. We observed the creation of many period-1 UPOs, and as the parameter $b$ was increased more of them emerge in the interval $a \in [0,1]$. We computed numerically the global bifurcation diagrams, which agree with the analytical results. The diagrams also reveal periodic orbits with higher periods, which are created and bifurcate similarly as the period-1 UPOs. We argue that the large number of UPOs are responsible for the creation of the global chaotic saddle (GCS).

In the bifurcation diagrams we observed the shearless attractor (SA), predominantly chaotic and with some windows of periodicity. We analyzed a periodic window that begins when the chaotic SA loses its stability due to the birth of a period-26 attractor, via a saddle-node bifurcation. At this moment the LCS is created. We showed that the presence of the second chaotic saddle increases the transient times. The period-26 attractor goes through a period-doubling cascade creating a small banded chaotic attractor. This banded attractor collides with the LCS causing a local interior crisis giving rise to an enlarged single-banded attractor.

We observed a second interior crisis as well, that happens when the chaotic SA collides with the GCS, resulting in an attractor that spreads itself in a large range of the y-coordinate. Both crises produce bursts of intermittency, and we studied the intermittent behavior induced by the global interior crisis. We found that the average time between consecutive bursts decreases quickly as the control parameter $a$ is increased.

To conclude we make a few remarks about generality. For the computations here we chose the parameter value $\gamma = 0.1$. The limit of $\gamma$ approaching zero is of interest. For $\gamma$ at zero, we have the well-known conservative standard nontwist map, which has been most thoroughly investigated in the literature. As gamma nears zero, however, the following happens: some of the stable periodic orbits become periodic attractors and the smaller gamma becomes, the more of these periodic attractors are present. Also, if there is a shearless curve for $\gamma = 0$, it limits to a quasiperiodic attractor as the dissipation is added. As $\gamma$ gets smaller, the transients seem to be longer as it takes more iterations for an initial condition to reach an attractor. Although we have observed these results for small $\gamma$, we have not done a systematic study of this limit. The limit of strong dissipation, where $\gamma$ approaches 1, results in a one-dimensionnal circle map in the x-variable. We are currently investigating this limiting case, and we already have several analytical and numerical results that we will report on in the future. We note, the choice of $\gamma = 0.1$ was made because this particular parameter value has been used in previous works (e.g., [26, 27]). The results presented in the manuscript are general and we have observed the same phenomena for other values of gamma between 0 and 1. Finally, the results here were obtained for a specific discrete time system, but they are expected to be general and observable in other dissipative systems, either continuous or discrete.

The necessary condition is to have multiple chaotic saddles allowing for qualitatively different interior crises.

**ACKNOWLEDGEMENTS**

*We acknowledge support from the Brazilian scientific agency CAPES- Coordination for the Improvement of Higher Education Personnel. REC also thanks FAPESP-São Paulo Research Foundation through the grant 2019/07329-4. RLV also thanks CNPQ-National Council for Scientific and Technological Development through the grants 403120/2021-1 and 301019/2019-3, CAPES through the grant 88881.143103/2017-01, and FAPESP through the grant 2022/04251-7. PJM was supported by the U.S. Department of Energy Contract N° DE-FG05-80ET-53088.*

**APPENDIX – Numerical Algorithms**

In this appendix we briefly review the numerical algorithms utilized throughout this study. Two popular algorithms for finding chaotic saddles are the sprinkler [41,42] and the proper interior maximum triple (PIM) [43] methods. We describe the implementation of these methods for two-dimensional maps, such as the DSNM, but they can be generalized to higher dimensional and continuous time systems. In our implementation it is necessary to identify when trajectories have reached an attractor, so we start by describing how to do so by means of the attractors natural measure [44].

1. **Natural measure**

The natural measure $\mu$ [44] quantifies how often different regions of the attractor are visited by a dense trajectory. It yields the probability that a point moving on the attractor for a long time will visit different regions. This measure is a characteristic of the attractor itself and therefore is not dependent on the initial condition, any starting point on the attractor's basin of attraction can be used to compute the natural measure.

Let $\phi(x_0, y_0)$ be a trajectory of the system with initial condition $(x_0, y_0)$. Assume that the point $(x_0, y_0)$ belongs to the basin of attraction of the attractor we desire to compute the natural measure. We cover a phase space region that contains the attractor with boxes of side-length $\epsilon = 10^{-4}$. Let $\eta(B_i, \phi(x_0, y_0), T)$ be the total number of iterations spent by the solution inside the box $B_i$, considering $T$ iterations of the map.

If $\eta$ is the same for almost every initial condition on the attractor's basin of attraction, then the natural measure of each box is defined as

$$\mu_i = \lim_{T \to \infty} \frac{\eta(B_i, \phi(x_0, y_0), T)}{T} \qquad (A1)$$

if the limit exists. By this definition, $\sum_{i=1}^{N} \mu_i = 1$, where $N$ is the total number of boxes. In our simulations we consider $T = 10^8$. Once the natural measure is obtained it is straightforward to verify if a trajectory has reached the attractor. The convergence to the attractor happens when a box $B_i$ with $\mu_i \neq 0$ is visited for the first time. In configurations with multistability, where multiple attractors coexist, it is necessary to evaluate the natural measure of each of them.

2. **Sprinkler method**

The idea of the sprinkler method [41,42] is to follow an ensemble of trajectories and select pieces of them that remain in the neighborhood of a chaotic saddle $\Lambda$. The stable manifold of $\Lambda$ is the set of points that converge to $\Lambda$ under forward iterations of the map, while the unstable manifold of $\Lambda$ is the set of points that converge to $\Lambda$ under backwards iterations of the map. The chaotic saddle itself is the intersection between its stable and unstable manifolds.

Consider a region $R$ in the phase space containing at least a part of a chaotic saddle and with no attractors inside of it; we distribute a large number $N_0 = 10^7$ of initial conditions in this region and iterate them under the map. All points in $R$ eventually converge to an attractor, except points lying on the stable manifold, which is a measure zero set. The initial conditions close to the stable manifold are attracted to Λ and stays on its neighborhood for some time, and then are repelled by the unstable manifold towards an attractor. The sprinkler algorithm relies on those facts and can approximate the chaotic saddle Λ and its stable and unstable manifolds.

We calculate the number of iterations necessary for each initial condition to reach an attractor and discard the ones with escape times shorter than a specified time $n_c$. The trajectories with escape times larger than $n_c$ come close to Λ during the dynamical evolution, implying that the corresponding initial conditions approximate the stable manifold of Λ. On the other hand, their last iteration before reaching the attractor approximate the unstable manifold. The chaotic saddle is approximated by the points at the middle of these trajectories, obtained after $\xi \approx n_c/2$ iterations. In summary, the first, middle and last points of the trajectories with large escape times approximate the stable manifold, the chaotic saddle, and the unstable manifold, respectively. The time $n_c$ is chosen by trial and error but must be larger than the average escape time $\tau$ defined by Eq. (20). The sprinkler method has been widely used due to its simplicity in implementation, low computational cost and the ability to obtain the manifolds of the chaotic saddle.

## 3. PIM triple method

While the sprinkler method approximates a chaotic saddle with fragments of several different trajectories, the PIM triple method [43] aims to find a single trajectory that stays close to Λ for an arbitrarily long time. Just as the sprinkler method, the PIM triple relies on the observation that trajectories starting close to the stable manifold of the saddle stay for a long time in the vicinity of the saddle.

The algorithm starts with two points $a$ and $b$ such that the interval $[a, b]$ is a line segment that intersects the stable manifold of Λ. Then there will be a point $c \in [a, b]$ that is closer to the saddle's stable manifold than $a$ and $b$; therefore, the escape time of $c$ is larger than the escape time of $a$ and $b$. A refining of the initial line segment is done as follows: we uniformly distribute 1000 points on the interval $[a, b]$, measure their escape times and identify the triple $(a_1, c_1, b_1)$ of consecutive points, such that $c_1$ is the point with larger escape time in the interval $[a, b]$.

The triple $(a_1, c_1, b_1)$ is called an *Interior Maximum* triple, and the new, smaller, line segment $[a_1, b_1]$ also crosses the stable manifold. The refining process is repeated generating new intervals $[a_2, b_2]$, $[a_3, b_3]$, and so on, until we find a line segment $[a_n, b_n]$ with length smaller than $\delta = 10^{-10}$. The triple $(a_n, c_n, b_n)$, obtained after the refinement, is called a *Proper Interior Maximum* (PIM) triple.

Let $I_0$ be the interval $[a_n, b_n]$. Then the triple $(a_n, c_n, b_n)$ is iterated under the map, obtaining a new interval $I_1$. If the length of $I_1$ is smaller than $\delta$, it is iterated further until its length becomes greater than $\delta$, and in that case the refining procedure is restarted to obtain a new interval $I < \delta$. The algorithm proceeds to iterate $I$ and doing the refining procedure every time the length of the interval exceeds $\delta$.

With the described method we find a set of PIM triples with size smaller than $\delta$, and the set of middle points approximates a typical trajectory on the chaotic saddle, called the straddle trajectory, which stays within a distance of the order of $\delta$ from the actual chaotic saddle.

Although the PIM triple method is computationally expensive when compared to the sprinkler method, it provides a more accurate approximation of the chaotic saddle. All chaotic saddles presented in this work were obtained with the PIM triple method and $\delta = 10^{-10}$.